\documentclass[aps,prl,letterpaper,amsmath,amssymb,superscriptaddress,reprint,
]{revtex4-1}
\usepackage{graphicx}

\begin{document}
\title{Experimental realization of a semiconducting\\full Heusler compound: Fe$_2$TiSi}
\author{Markus Meinert}
\email{meinert@physik.uni-bielefeld.de}
\author{Manuel P. Geisler}
\author{Jan Schmalhorst}
\affiliation{Center for Spinelectronic Materials and Devices, Bielefeld University, D-33501 Bielefeld, Germany}
\author{Ulrich Heinzmann}
\affiliation{Molecular and Surface Physics, Department of Physics, Bielefeld University, D-33501 Bielefeld, Germany}
\author{Elke Arenholz}
\affiliation{Advanced Light Source, Lawrence Berkeley National Laboratory, CA 94720, USA}
\author{Walid Hetaba}
\affiliation{Center for Spinelectronic Materials and Devices, Bielefeld University, D-33501 Bielefeld, Germany}
\affiliation{University Service Center for Transmission Electron Microscopy, Vienna University of Technology, A-1040 Vienna, Austria}
\author{Michael St\"oger-Pollach}
\affiliation{University Service Center for Transmission Electron Microscopy, Vienna University of Technology, A-1040 Vienna, Austria}
\author{Andreas H\"utten}
\author{G\"unter Reiss}
\affiliation{Center for Spinelectronic Materials and Devices, Bielefeld University, D-33501 Bielefeld, Germany}
\date{\today}

\begin{abstract}
Single-phase films of the full Heusler compound Fe$_2$TiSi have been prepared by magnetron sputtering. The compound is found to be a semiconductor with a gap of 0.4\,eV. The electrical resistivity has a logarithmic temperature dependence up to room temperature due to Kondo scattering of a dilute free electron gas off superparamagnetic impurities. The origin of the electron gas is extrinsic due to disorder or off-stoichiometry.  Density functional theory calculations of the electronic structure are in excellent agreement with electron energy loss, optical, and x-ray absorption experiments. Fe$_2$TiSi may find applications as a thermoelectric material.
\end{abstract}

\maketitle
Energy harvesting has become an important technology with large industrial and economic impact \cite{Vullers09}. Among the various harvesting principles, thermoelectric power generation is particularly useful for devices worn close to the human body (like wrist watches) or for sensors in industrial process monitoring. The key ingredient in such a device is the thermoelectric material, which usually is a doped semiconductor. Finding materials with a high thermoelectric efficiency is nowadays an important challenge. Among the class of half-Heusler compounds (C1$_\mathrm{b}$ structure) many semiconducting materials with promising thermoelectric properties are known \cite{Graf11,Casper12,Bartholome13}. In contrast, semiconducting full-Heusler compounds (L2$_1$ structure, called simply Heusler compounds in the following) appear to be rare; in fact, there is no clear experimental evidence for a semiconducting ground state with a notable energy gap in any Heusler compound, despite 110 years of research on this material class \cite{Heusler1903}. Still, some Heusler compounds are known to be semimetals, among which Fe$_2$VAl is the most prominent example \cite{Nishino97, Okamura00}. For Si- or Ge-doped Fe$_2$VAl, interesting thermoelectric properties were reported \cite{Nishino06,Lue07}.

It is generally understood, that only a Heusler compound with 24 valence electrons can be a paramagnetic semiconductor \cite{Galanakis02}. The Heusler compound Fe$_2$TiSi fulfills this condition and has recently been predicted to be a semiconductor with a sizeable gap (0.41\,eV) and large Seebeck coefficient ($-300\,\mu$V/K at room temperature) \cite{Yabuuchi13}. The synthesis of this compound has been attempted more than 40 years ago, however it was found to be metastable and to concur with Fe$_2$Ti in the hexagonal C14 Laves Phase \cite{Jack72, Niculescu77}. The closely related Fe$_2$TiSn is found to be a semimetal \cite{Slebarski00, Dordevic02}.

In this article we describe the synthesis of single-phase Fe$_2$TiSi thin films with a highly ordered full Heusler (L2$_1$) structure, which is stabilized by epitaxy. This allows for studying the electronic structure of this unique material and for performing a detailed comparison with calculations. Epitaxial thin films of Fe$_2$TiSi with thicknesses of $22$\,nm and $110$\,nm were grown on MgAl$_2$O$_4$(001) substrates by dc and rf magnetron co-sputtering with a substrate temperature of up to 780$^\circ$C and a growth rate of $0.1$\,nm/s. Elemental Fe, Ti, and Si targets with a higher purity than 99.99\%  were used. The films were capped with $3$\,nm Si to prevent them from oxidation.

\begin{figure}[t]
\includegraphics[width=8.6cm]{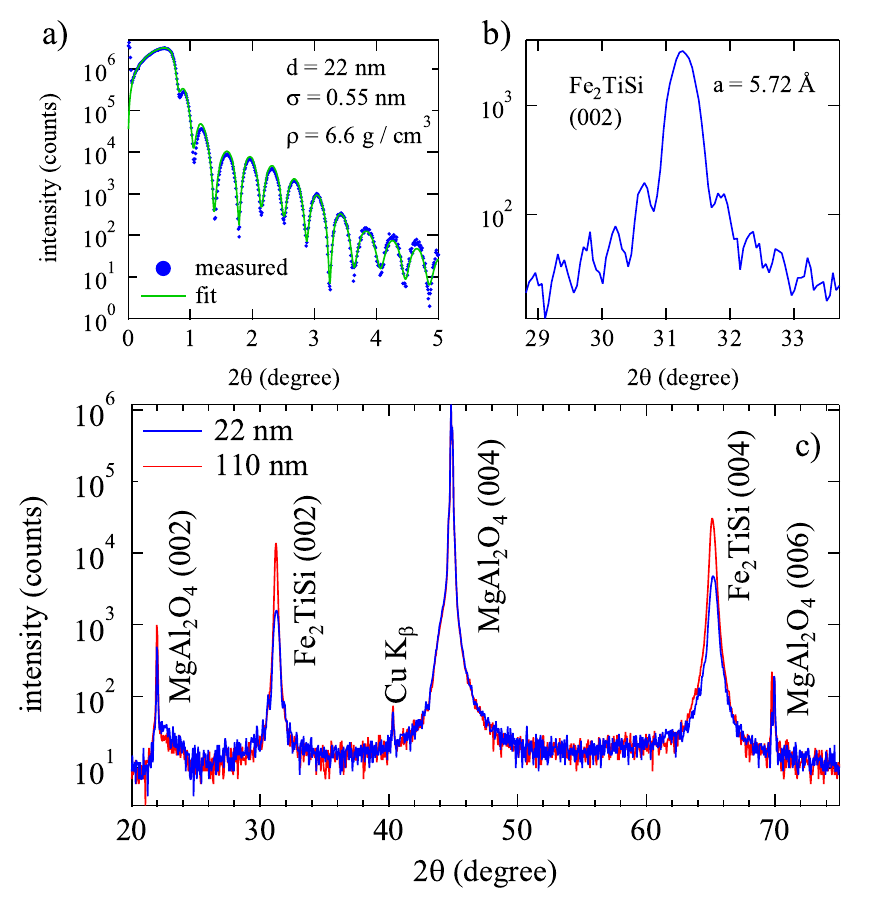}
\caption{\label{xrd}a): X-ray reflectivity spectrum with fit for a 22\,nm film. b): High resolution scan of the (002) peak of a 22\,nm film, which exhibits Laue oscillations. c): X-ray diffraction spectra of 22\,nm and 110\,nm thick films. All visible peaks belong to either the films or the substrates..}
\end{figure}

Film compositions were determined by x-ray fluorescence spectroscopy with an accuracy of typically $\pm 1\%$. Film thicknesses and roughnesses, order parameters, and lattice constants were obtained by x-ray reflection and diffraction with Cu K$_\alpha$ radiation. The sputtered films have smooth surfaces and exhibit Laue oscillations on their diffraction peaks (Fig.~\ref{xrd} a) and b)), which are indicative of coherent crystal growth, i.e., the crystal lattice planes are strictly parallel to the film interfaces. The x-ray reflectivity fit reveals a standard deviation of the nominal thickness (a measure for the roughness) of $\sigma = 0.55\,\mathrm{nm}$ and a density of $(6.6 \pm 0.05)\,\mathrm{g/cm}^3$. From x-ray diffraction we obtain the lattice constant $a = 5.72$\,\AA{}. The theoretical value of $a = 5.717$\,\AA{} (bulk modulus $B_0 = 235$\,GPa) obtained with the \textsc{elk} density functional theory (DFT) code \cite{elk} using the Perdew-Burke-Ernzerhof (PBE) functional \cite{PBE} matches the experimental value very well. The calculated density of the material is $6.66\,\mathrm{g/cm}^3$, in excellent agreement with the measured value. The lattice constant matches the lattice constant of the substrates within $0.1\,\%$ with the epitaxial relation Fe$_2$TiSi [110] $||$ MgAl$_2$O$_4$ [100]. This allows for nearly unstrained growth, which is confirmed by the film thickness limited peak widths and the rocking curve width of $0.4^\circ$. The x-ray diffraction measurements (Figure \ref{xrd} c)) show no peaks that could be assigned to other than the Heusler phase of Fe$_2$TiSi. By comparing the observed with the peak intensity ratios calculated for the perfectly ordered compound we determine the order parameters $S_\mathrm{B2}$ and $S_\mathrm{L2_1}$ as a measure of the chemical disorder in our films \cite{Takamura09}. The calculations include anomalous corrections and a Debye model for the thermal motion of the atoms, for which we assume a Debye temperature of 450\,K \cite{Kanchana09}. The exact choice of this value is uncritical, though. We find $S_\mathrm{B2} = 0.99 \pm 0.02$ and $S_\mathrm{L2_1} = 0.98 \pm 0.05$ for the best films, which were deposited at 780$^\circ$C on MgAl$_2$O$_4$ and will be discussed in the following. These values are indicative of a very good chemical order, although Fe-Ti disorder is hard to detect. Site-swapping or off-stoichiometry may be present in our films in the range of a few percent, according to the experimental uncertainties in the XRD and XRF measurements. Scanning electron microscopic images (not shown) further confirmed the homogeneity of the films. Thus, we conclude that we have grown single phase Fe$_2$TiSi with the full Heusler structure.

\begin{figure}[b]
\includegraphics[width=8.6cm]{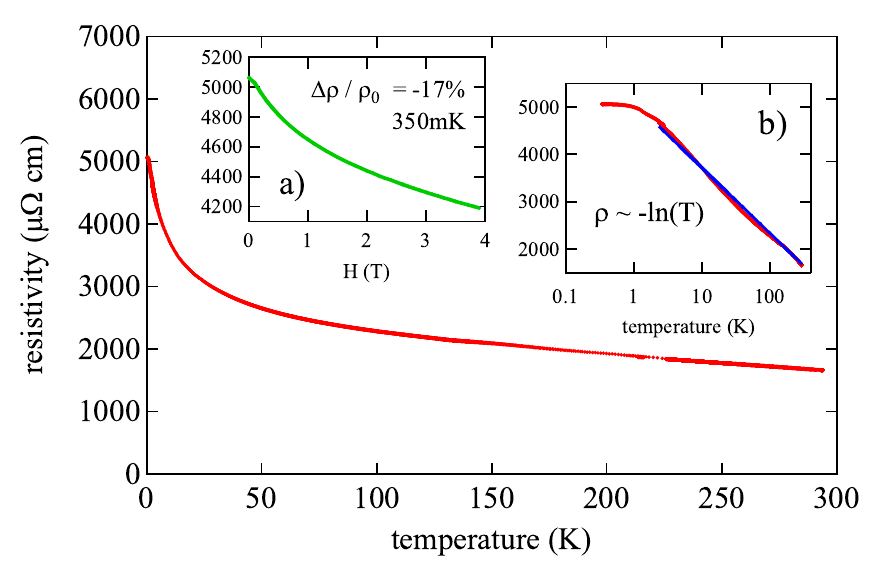}
\caption{\label{resistivity} Resistivity of a 110\,nm thick film of Fe$_2$TiSi as a function of temperature. The negative magnetoresistance (inset a)) and the logarithmic temperature dependence of the resistivity (blue line in inset b)) indicate Kondo scattering.}
\end{figure}

Temperature dependent resistivity measurements in a $^3$He dilution cryostat were taken down to 0.35\,K in magnetic fields up to 4\,T. We find a negative temperature coefficient of the resistivity and an essentially logarithmic temperature dependence above 2\,K. This is shown for a 110\,nm thick film in Fig.~\ref{resistivity} and in its inset b). The logarithmic temperature dependence (fit in inset b)) indicates Kondo scattering as the main mechanism that causes the electrical resistivity \cite{Kondo64}. This is corroborated by the finite resistivity for $T \rightarrow 0$\,K and a negative magnetoresistance of $-17\,\%$ at 4\,T and 350\,mK \cite{Kondo64, Andrei82}, see inset a) in Fig.~\ref{resistivity}. The dependence of the resistivity on the external magnetic field agrees with that of other Kondo systems \cite{Giordano96}. The effective carrier concentration obtained from the Hall effect is $n=4.05 \cdot 10^{20}\,\mathrm{cm}^{-3} = 0.019\,/\,\mathrm{f.u.}$ at room temperature and increases to $n=5.50 \cdot 10^{20}\,\mathrm{cm}^{-3}$ at 2\,K. This indicates that electron- and hole-like trajectories with different mobilities and different temperature dependencies contribute to the conductivity. An anomalous contribution to the Hall effect is observed, which points to superparamagnetic order. The scattering time is calculated as $\tau(\mathrm{RT}) = 5.27 \cdot 10^{-15}$\,s and $\tau(2\,\mathrm{K}) = 1.48 \cdot 10^{-15}$\,s by assuming the effective mass to be $m^* = m_e$. The Kondo effect being the dominating mechanism for the resistivity up to room temperature is uncommon, but not unheard of \cite{Schoenes87}.

\begin{figure}[t]
\includegraphics[width=8.6cm]{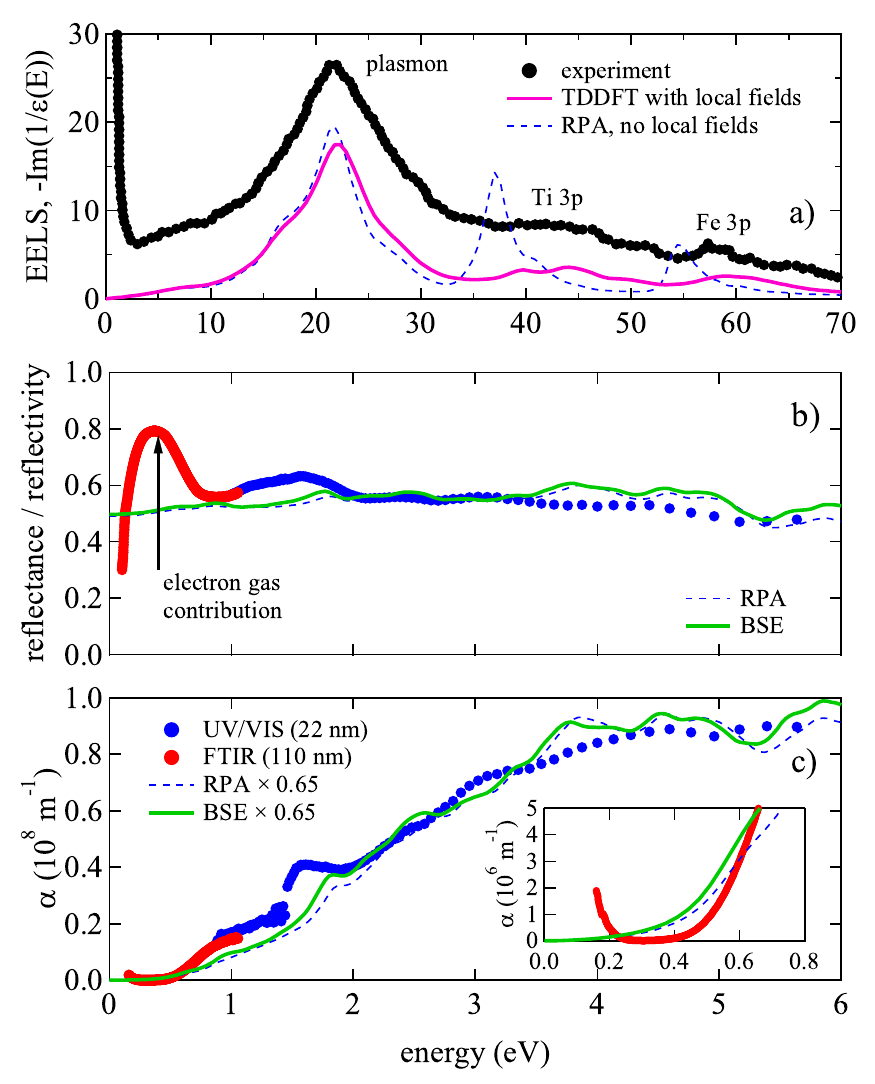}
\caption{\label{eels_optics}a): Measured and calculated electron energy loss spectra. b): Measured reflectance and calculated reflectivity. c): Measured and calculated absorption coefficient. The measurements were done on films with two different thicknesses to improve the signal-to-noise ratio.}
\end{figure}

Valence band electron energy loss spectroscopy (EELS) up to 70\,eV was taken in a FEI TECNAI G20 at 20\,kV accelerating voltage with the scattering vector $\mathbf{q} \rightarrow 0$. Optical absorption and reflectivity measurements were done with a Perkin Elmer Lambda 950 UV/Vis spectrometer and a Thermo Scientific Nicolet 8700 Fourier transform infrared (FTIR) spectrometer for photon energies between 0.1 and 5.5\,eV at room temperature. The results are given in Fig.~\ref{eels_optics}. Two different thicknesses (22\,nm and 110\,nm) were used to optimize the signal-to-noise ratio for the different energy regimes of the two spectrometers. The optical absorption coefficient $\alpha$ has been obtained from the reflectance $R$ and the transmittance $T$ as \cite{Cantarero88}
\begin{equation}
\alpha = \frac{1}{d} \ln \left( \frac{(1-R)^2}{2T} + \sqrt{\left( \frac{(1-R)^2}{2T} \right)^2 + R^2} \right),
\end{equation}
where $d$ is the film thickness. The absorption spectrum shows a gap below about 0.4\,eV (see inset in Fig.~\ref{eels_optics} c)). The reflectance spectrum increases strongly at photon energies below 0.75\,eV, which is the free-electron plasma frequency based on the carrier density obtained from the Hall effect. Thus, the large low-energy reflectance and the increase of absorption below 0.25\,eV originate from the free carriers with low density. The sharp drop of the reflectance below 0.4\,eV arises from the fact that the radiation of wavelength $\lambda$ penetrates through the thin film when $\lambda \gg d$.

For comparison, the electronic structure was calculated with the \textsc{elk} code. Optical spectra were calculated within the random phase approximation (RPA) and by solving the Bethe-Salpeter-Equation (BSE) \cite{Rohlfing00}, for which the Brillouin zone was sampled with a shifted $8 \times 8 \times 8$ $\mathbf{k}$-point mesh. A Lorentzian broadening of 0.1\,eV was applied. The loss function $-\mathrm{Im}(1/\varepsilon(q=0,E))$ was calculated with time-dependent DFT (TDDFT) in the adiabatic approximation including local field effects (LFE) with a broadening of 1\,eV. The EELS is dominated by a  plasmon resonance at 21.6\,eV, see Fig.~\ref{eels_optics}~a). The free-electron theory predicts 26.6\,eV, thus the average effective mass of the valence electrons has to be 1.52\,$m_\mathrm{e}$ to match the experimental value. The plasmon energy in the computed spectra matches the experimental value exactly. The general accuracy of computed loss functions based on DFT calculations is well-known \cite{Keast05}. For the Ti and Fe 3$p$ semicore excitations inclusion of LFE is mandatory \cite{Vast02} and doing so leads to a quantitative agreement with the experiment. The calculated optical reflectivity and absorption spectra are overall in reasonable agreement with the measured data. Both the computed and the experimental absorption spectra show the gap at low energy, although the experimental gap is about 0.1\,eV larger, as expected for a standard DFT calculation. The pronounced features around 1.6\,eV in the experimental spectra are also found in the computed spectra, although somewhat weaker and at higher energy. The BSE spectrum indicates an excitonic enhancement in this region of the spectrum, however not enough to match the experiment. A frequency-dependent screening in the BSE might be necessary to capture this feature properly as well as the absolute value of the absorption coefficient  \cite{Marini03}.  Excitonic effects are overall weak due to the large dielectric constant $\varepsilon \approx 35$. For a good agreement of the calculated optical properties with the experimental data, a good description of both occupied and unoccupied bands and corresponding wavefunctions is necessary. Thus, we conclude that the electronic structure of Fe$_2$TiSi is overall well described by the DFT calculation.

\begin{figure}[b]
\includegraphics[width=8.6cm]{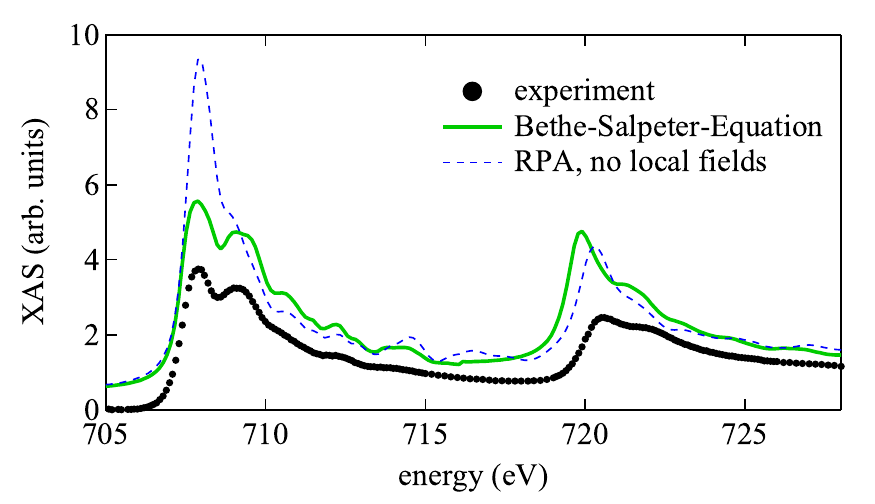}
\caption{\label{xas} Experimental and computed x-ray absorption spectra of Fe in Fe$_2$TiSi. The spectra are scaled to one at 40\,eV above the absorption onset.
}
\end{figure}

\begin{figure}[b]
\includegraphics[width=8.6cm]{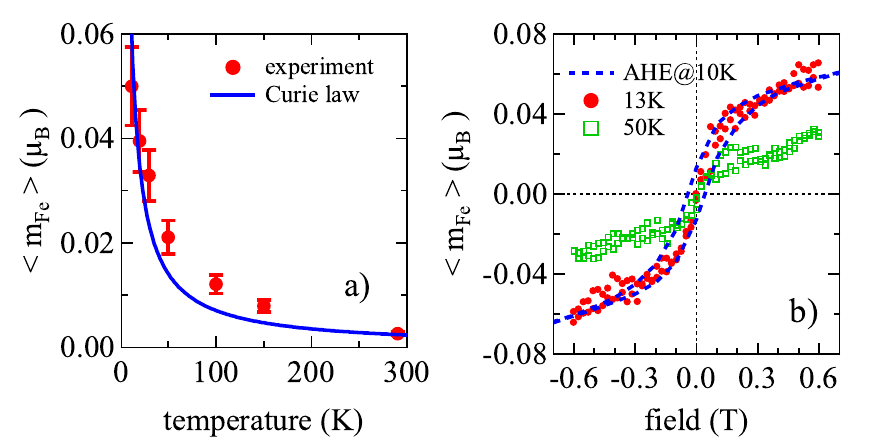}
\caption{\label{xas_magnetism}a): Temperature dependence of the average magnetic moment of Fe at 0.5\,T with a fit to the Curie law. b): Magnetic loops perpendicular to the film plane taken on Fe and anomalous Hall effect (AHE) at 10\,K for comparison.}
\end{figure}

The structure of the unoccupied Fe states was investigated with x-ray absorption spectroscopy at the Fe $L_{2,3}$ edges. The x-ray absorption (XAS) and magnetic circular dichroism (XMCD) spectra were taken at temperatures down to 13\,K at beamline 4.0.2 of the Advanced Light Source, Berkeley. The substrate luminescence was detected with a photodiode in addition to the total electron yield to measure the absorption signal of the films. The experimental spectrum in Fig.~\ref{xas} shows a strong double-peak structure and some weaker features at the high-energy side of the absorption edge. X-ray absorption spectra were calculated in the random phase approximation without local fields and by solving the Bethe-Salpeter-Equation. Here, the Brillouin zone was sampled with a shifted $6 \times 6 \times 6$ $\mathbf{k}$-point mesh and spin-orbit coupling was included. The exchange term of the BSE Hamiltonian was scaled to 85\,\% as suggested by Vinson \textit{et al.} \cite{Vinson12}. A Lorentzian broadening of 0.27\,eV (0.44\,eV) was applied to the L$_3$ (L$_2$) edges to account for their different core-hole lifetimes. Recent studies have found the BSE to be remarkably successful in reproducing experimental L$_{3,2}$ absorption spectra of 3\textit{d} transition metals and their compounds \cite{Laskowski10, Vinson12}. The calculated absorption signal in the random phase approximation is dominated by a strong peak at the absorption onset and some shoulders at higher energy, at odds with experiment. Explicitly treating the electron-hole interaction with the Bethe-Salpeter-Equation causes a redistribution of spectral weight from the main peak into the higher-energy shoulders and gives satisfactory agreement between measured and calculated $L_3$ spectra. The main double-peak structure stems from the $e_g - t_{2g}$ splitting of the Fe states, which are mixed due to the electron-hole interaction \cite{Laskowski10}. Notably, this structure is also found in Co$_2$TiSn \cite{Meinert11} and indicates the hybridization between the atoms at the Wyckoff 8c (Co or Fe) positions with the high-lying $d$-states of Ti as the driving mechanism for the large crystal field splitting. However, the branching ratio and the spin-orbit splitting between $L_3$ and $L_2$ are too small in the BSE spectrum. This originates from the applied approximation for the wave functions: the relativistic Fe $2p$ states are described with a single scalar-relativistic local-orbital and spin-orbit coupling is added in second variation. Thus, the mixing between the well described $2p_{3/2}$ and the improperly described $2p_{1/2}$ transitions is incorrect. For the same reason, the $L_3$ edge is overall in better agreement with the experiment than the $L_2$ edge.

\begin{figure}[t]
\includegraphics[width=8.6cm]{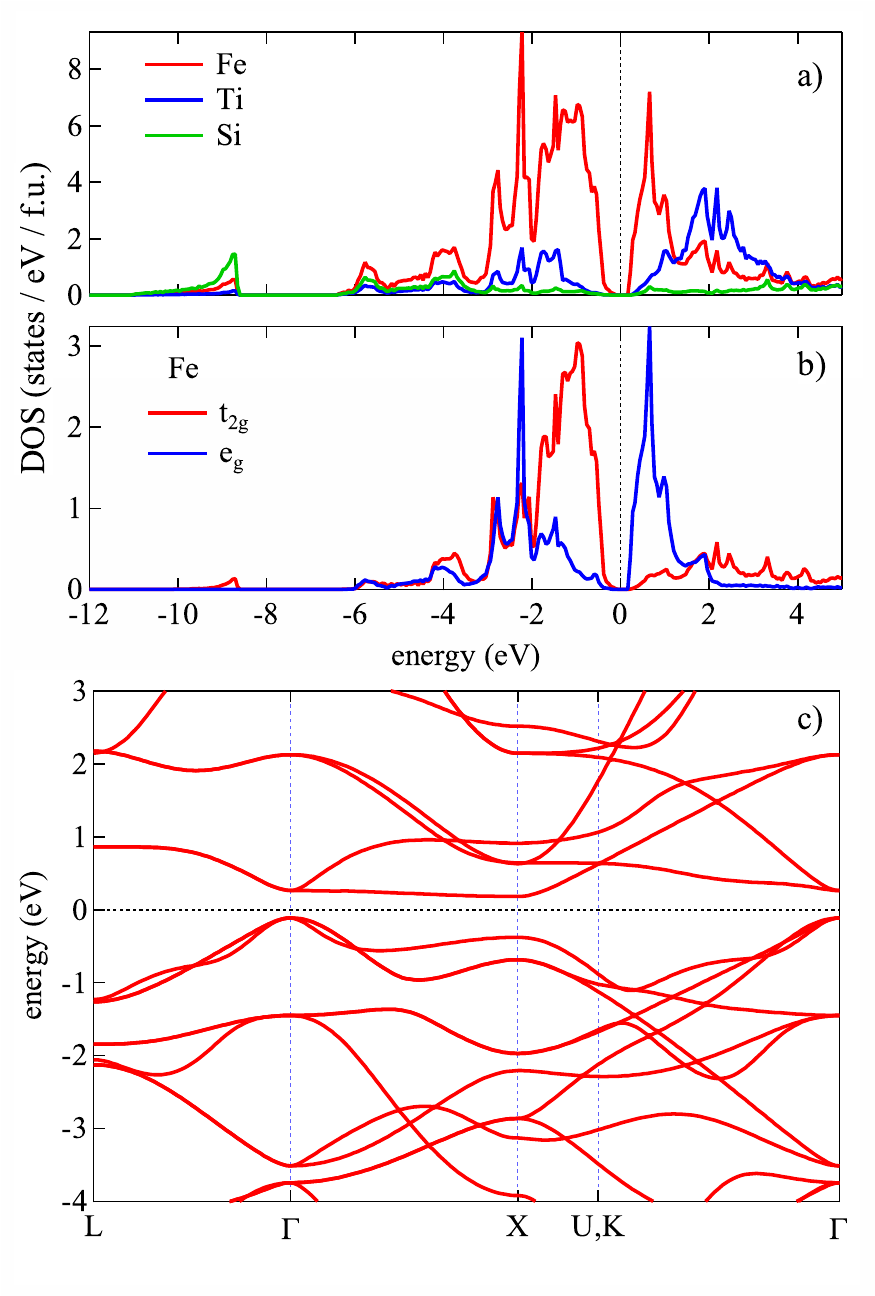}
\caption{\label{band}a): Element resolved density of states plots of Fe$_2$TiSi. b): Site-symmetry resolved density of states of Fe in Fe$_2$TiSi. The $e_g - t_{2g}$ crystal field splitting is responsible for the band gap formation. c): Band structure plot of Fe$_2$TiSi.}
\end{figure}

The characteristic double-peak structure in the XAS results of the crystalline films is absent for amorphous samples and it appears as a consequence of crystallization.  The amorphous film is ferromagnetic and has a magnetic moment of $0.6\,\mu_\mathrm{B}$\,/\,Fe at 20\,K. This moment nearly vanishes for the crystallized material. Still, the XMCD of the crystalline material indicates weak magnetism carried by Fe atoms, see Fig. \ref{xas_magnetism}. The temperature dependence of the average Fe moment $\left< m_\mathrm{Fe} \right>$ is similar to a paramagnetic dependence, however, the experimental data deviate somewhat from the Curie law (Fig. \ref{xas_magnetism} a)). The magnetic loops taken on Fe indicate that not only a paramagnetic but also a superparamagnetic or very weakly ferromagnetic component are present, very similar to the case of Fe$_2$VAl \cite{Lue01}. Remarkably, the anomalous Hall effect shows a nonzero coercivity, which is not observed in the XMCD loops (Fig. \ref{xas_magnetism} b)). The presence of magnetic impurities is expected, as some Fe atoms may occupy antisite positions, on which they are expected to exhibit a large localized magnetic moment \cite{Singh98}. The impurities may form superparamagnetic clusters, which give rise to the observed characteristic magnetic loops \cite{Feng01}.

Because the DFT calculation provides a good description of the electronic structure of Fe$_2$TiSi, we discuss the theoretical results in more detail. The density of states (DOS) for Fe, Ti, and Si is given in Fig.~\ref{band} a). It becomes clear, that the edges of the band gap are entirely defined by Fe states. This is in contrast to Fe$_2$VAl, in which V contributes significantly to the formation of the pseudogap \cite{Singh98}. In Fe$_2$TiSi, the unoccupied Ti states are well separated from the Fe states, which allows to form a real band gap. Investigating the Fe states in terms of their symmetry, we find that the band gap arises from the crystal field splitting into doubly and triply degenerate $e_g$ and $t_{2g}$ states, which define the band gap (Fig.~\ref{band} b)). As shown in Fig.~\ref{xas}, this splitting is directly observed in the Fe XAS. The same mechanism is responsible for the formation of the gap in half-metallic Heusler compounds \cite{Galanakis02}. The band plot (Fig.~\ref{band} c)) reveals that the band gap is of indirect nature from $\Gamma$ to $X$. However, the minimum direct gap at $\Gamma$ is just 0.08\,eV larger, so indirect transitions may be expected to play no significant role in the optical absorption. The large hybridization gap between the Si $s$ states below $-8$\,eV and the higher states gives rise to the highly ordered structure of Fe$_2$TiSi thanks to a strongly covalent character of the Si bonds. This is in contrast to Fe$_2$VAl \cite{Singh98}, which is very susceptible to V-Al site swap disorder \cite{Feng01}.

In summary, we have prepared single phase films of the metastable Fe$_2$TiSi full Heusler compound and studied their electronic and transport properties in detail. The most important finding is that a band gap of about 0.4\,eV is present, as predicted by electronic structure theory. This constitutes the first experimental report of a semiconducting full Heusler compound. Arising from residual off-stoichiometry and disorder, the electrical transport properties are governed by a small density of quasi free electrons that scatter off magnetic impurities. Remarkably, this Kondo scattering behaviour is observed up to room temperature. Future studies may be directed towards the proposed thermoelectric applications \cite{Yabuuchi13} and towards a possible connection between the Seebeck effect and disorder-induced magnetic properties. Replacing some Si with Sn will simultaneously reduce the electrical resistivity and the lattice thermal conductivity, thereby increasing the thermoelectric power.

We thank the Ministerium f\"ur Innovation, Wissenschaft und Forschung des Landes Nordrhein-Westfalen (MIWF NRW) for financial support. We also thank the developers of the \textsc{elk} code, for their efforts. The Advanced Light Source is supported by the Director, Office of Science, Office of Basic Energy Sciences, of the U.S. Department of Energy under Contract No. DE-AC02-05CH11231.

\end{document}